**Title of the article:**

Ethically Aligned Design: An empirical evaluation of the RESOLVEDD-strategy in Software and Systems development context

**Authors:**

Ville Vakkuri, Kai-Kristian Kemell, Pekka Abrahamsson

**Notes:**

- This is the author's version of the work.
- The definite version was published in: Vakkuri V., Kemell KK., Abrahamsson P. (2019) Ethically Aligned Design: An Empirical Evaluation of the RESOLVEDD-Strategy in Software and Systems Development Context, in the proceedings of the 45th Euromicro Conference on Software Engineering and Advanced Applications (SEAA), pp. 46-50, IEEE
- Copyright owner's version can be accessed at https://doi.org/10.1109/SEAA.2019.00015

# Ethically Aligned Design: An empirical evaluation of the RESOLVEDD-strategy in Software and Systems development context

Ville Vakkuri, Kai-Kristian Kemell, Pekka Abrahamsson

Faculty of Information Technology, University of Jyväskylä
Jyväskylä, Finland
ville.vakkuri@jyu.fi, kai-kristian.o.kemell@jyu.fi, pekka.abrahamsson@jyu.fi

*Abstract*—Use of artificial intelligence (AI) in human contexts calls for ethical considerations for the design and development of AI-based systems. However, little knowledge currently exists on how to provide useful and tangible tools that could help software developers and designers implement ethical considerations into practice. In this paper, we empirically evaluate a method that enables ethically aligned design in a decision-making process. Though this method, titled the RESOLVEDD strategy, originates from the field of business ethics, it is being applied in other fields as well. We tested the RESOLVEDD strategy in a multiple case study of five student projects where the use of ethical tools was given as one of the design requirements. A key finding from the study indicates that simply the presence of an ethical tool has an effect on ethical consideration, creating more responsibility even in instances where the use of the tool is not intrinsically motivated.

*Keywords*—*artificial intelligence, ethics, design methods, ethical tool, RESOLVEDD, developer commitment*

## 1. Introduction

Artificial Intelligence and Autonomous Systems (AI/AS) are becoming increasingly ubiquitous. No longer are robots only found in factories, working highly repetitive conveyor belt tasks in closed environments. With autonomous vehicles entering the roads and AI systems filtering job applications out on the field, AI/AS are growing increasingly influential on a societal scale. It is practically impossible to opt out of using AI systems, with e.g. AI-based surveillance systems tracking you regardless of your consent. Similarly, due to the cyber-physical nature of many AI systems, their damage potential is not as narrow or predictable as that of conventional, purely digital software systems.

The pervasiveness of AI/AS systems forces us to analyze more profoundly under what type of ethical norms, rules and regulations AI systems should operate, and what kind of ethical standards should designers and developers hold when building these systems. As software engineers, developers are constantly making decisions when building systems. In doing so, they build their own values into the systems, which end up reflecting their views [1]. It is known that developers are not well- informed and aware of ethics[2]. Combined with the current lack of tools to support ethical AI development, this results in a situation where developers do not have the necessary means to tackle potential ethical issues, or even recognize them during development. Ethical issues are often simplified or simply neglected, only to be re-discovered later during the operational life of these systems once the damage has already been done.

One solution to this problem is to offer the developers an ethical instrument or tool to support ethical considerations in design and value alignment. However, our understanding of what kind of methods should be used in introducing developers to ethics and how these proposed methods work in practice is lacking. Developers prefer simple and practical methods if they use methods at all [3]. Ultimately, ethics are currently not considered important by developers, and therefore tools for supporting ethical consideration should not be resource-intensive to adopt, lest developers potentially see them as a nuisance.

To begin tackling this issue, we tested an ethical tool from business ethics, the RESOLVEDD-strategy, in the context of AI/AS design. We conducted a multiple case study of five different prototype projects where the use of ethical tool was given as one of the design requirements for the teams. The goal of this study is to better understand how the introduction of an ethical tool affects developers' ethical consideration in the design process and how the RESOLVEDD-strategy works in the given context.

*A. Ethically Aligned Design*

*Ethically Aligned Design* [4] refers to the involvement of decision-making in practice and ethical consideration in a the practice and design AI and autonomous systems and technologies. Involving ethical consideration into the context of software and interactive systems design has a history of more than 30 years. For example, Computer Ethics pioneer Bynum [5] introduced adapting human values in design before the rise of human values emphasizing the role of computer ethics. In response to ethical issues related to software and interactive systems development, Friedman [6] introduced a theoretically grounded Value Sensitive Design (VSD) approach and a method for the design of technology that accounts for human values in a principled, structured, and comprehensive manner throughout the design process [6,7]. Over the years, VSD has been tailored into various different branches of methods. For example, Davis and Nathan [8] further developed VSD by reinforcing its philosophical foundations. Wynsberghe [9] presented the Care Centered Value-Sensitive Design (CCVDS) for care robotics. Miller, Friedman, and Jancke [10] proposed Value Dams and Flows method to address values-oriented design tradeoffs. As a result, VSD has become a domain-agnostic general model for consideration of human values in the design, implementation, use, and evaluation of interactive systems [8].

To better incorporate human values into the design process of AI systems, some AI-specific values have been proposed. For example, the importance of transparency in AI systems was emphasized by Bryson and Winfield [11]. Dignum [12] presented two more values in addition to transparency by presenting the ART principles (Accountability, Responsibility, Transparency) to guide ethical development of AI systems [12]. Finally, fairness of AI systems and freedom from machine bias have also gained a significant role as core values expected from AI systems [13].

To direct the discussion on aligning ethics with system design, the IEEE Global Initiative on Ethics of Autonomous and Intelligent Systems was launched. The initiative was branded under a concept titled Ethically Aligned Design (EAD), a construct we discussed at the start of this section. The initiative aims to encourage practitioners to consider and prioritize ethics in the development of AI/AS. So far, the initiative has defined values and ethical principles that prioritize human well-being in a given cultural context. These guidelines and values have been published online, first in two versions for comments 2016-2018 and full release in 2019 EAD First Edition [4].

Arguably, the key audience of the EAD thinking should be the developers of the AI systems. AI development, much like conventional software development, is a cognitive activity [14] where humans play a significant role in deciding how the system behaves. Extant research has established that developers' interests are driven by work related concerns [15]. Concerns are the foundation of developer commitment development in his/her work. *Commitment* (discussed in detail in the next section) is important as it directs attention and helps in maintaining the chosen course of action [3, 15]. Should EAD practices become used by the developer, it should contribute to his work related concerns and help the developer to accomplish his or her tasks.

*B. The RESOLVEDD-strategy*

The step-by-step decision-making tool titled the RESOLVEDD- strategy was first introduced by Pfeiffer and Forsberg [16]. Originally, the RESOLVEDD-strategy was intended for teaching practical ethics to bachelor students. The method helps those who do not have prior knowledge of ethics or philosophy to evaluate ethical principles in practice. This aspect of the RESOLVEDD-strategy makes it particularly appealing for the field of Software Engineering (SE) where few curriculums have traditionally included studies in ethics or philosophy.

The RESOLVEDD-strategy is based on professional ethics and approaches ethics from the point of view of personal ethical problems in work contexts. It is not connected to any particular ethics theory and it does not enforce any set of values on its would-be users. Instead, RESOLVEDD is intended to support its users in taking into account ethical issues and tackling them through their own set of values or through an ethics theory of their choice. [16]

The strategy is presented as a series of nine concrete steps portraying the rational ethical decision-making process. By using the method, one is able to justify and explain the decision-making process leading up to whatever actions were ultimately taken. It is intended to help its users understand the ethical issues present in their work and encourages them to address them in their way of choosing, though nonetheless without compromising ethical principles. Though it originates from the field of business ethics, the method can also be utilized for tackling ethical issues outside the field of business. [17]

The nine steps of the RESOLVEDD-strategy can be seen as a process depiction in Figure 1. While utilizing resolved, however, these nine steps can be freely and flexibly modified to better suit each use context [16].

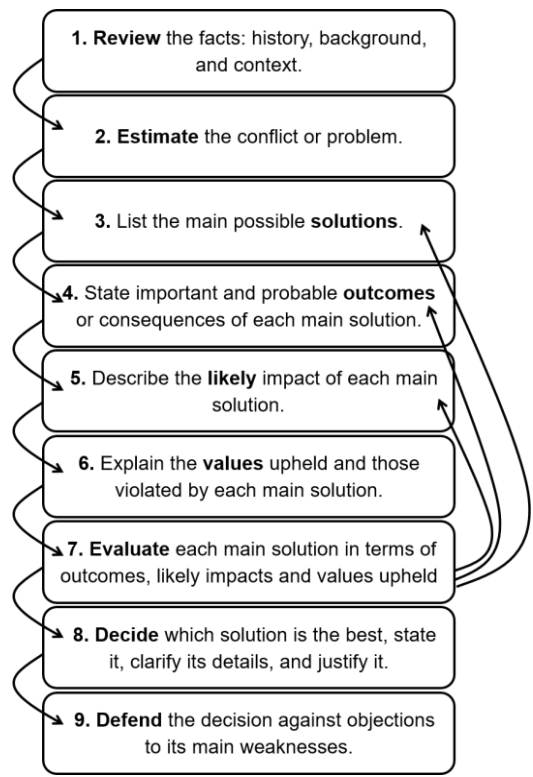

Figure 1. The Nine Steps of the RESOLVEDD-Strategy

## 2. RESEARCH FRAMEWORK AND STUDY DESIGN

*A. Research Framework*

In addressing ethical principles in AI/AS design, *accountability*, *responsibility,* and *transparency* (the ART principles) have recently been considered to be key constructs [12]. This study uses these three constructs as a basis and attempts to identify their possible relations, as well as relations of other constructs that may be involved in the process (Figure 2). The ART constructs have a central role in determining design protocols that take into consideration the designer, the product, and the end-users [12]. While other principles have been proposed for the ethical design of AI systems (see e.g. [4]), we consider the ART constructs a good starting point for understanding the involvement of ethics in ICT projects.

Developers' interests are driven by work-related concerns [15]. From the point of view of the developers, an important question to pose is: why would the developer act responsibly and take into account ethical issues? To begin tackling this question, meaningfulness of taken actions has been shown to be important in explaining work-related behavior [18]. For this reason, we need to understand the relationship between meaningfulness and the meaning of an activity, as we argue next. We have established that in order for an action to become meaningful for a developer, they must understand the meaning of the task. Therefore, a task that may be perceived as time consuming, boring, or otherwise lacking in motivational elements, will still be executed because it plays a role in the developer's commitment behavior [15].

Commitment, accountability, responsibility and transparency can therefore be seen as a cycle with links (Figure 2). These links are explorative as little empirical data is currently available. We can hypothesize that by strengthening commitment to the RESOLVEDD-strategy action, ethics will become implemented in the system. Ethics, as defined by EAD, is evidenced by increased in responsibility in design and clarity of accountability in order to help create more transparent culture in development of AI/AS. Transparent culture can likewise influence commitment, responsibility and accountability in design. In order to achieve this goal, the RESOLVEDD-strategy should (1) support responsibility, responsible culture, (2) help people to make more meaningful decisions in their own work, and (3) take into consideration ethical principles such as accountability, privacy, autonomy, and fairness.

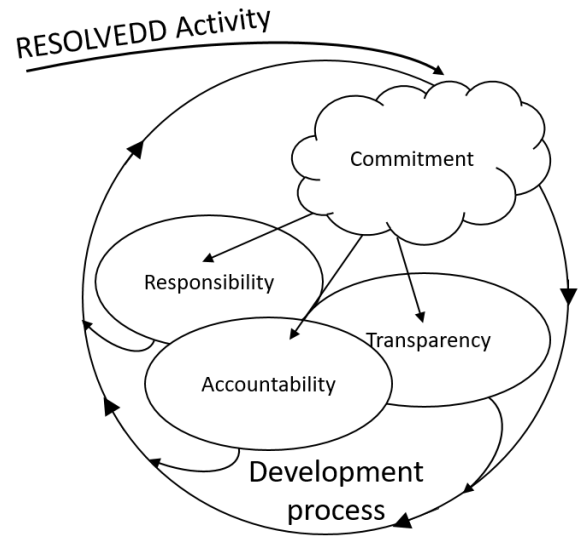

Figure 2. Framework for Ethically Aligned Design

*1) Commitment*

*Commitment* is the psychological bond between a person and an object (of the commitment) [19]. This bond is characterized by focus, strength, and type. The focus of the commitment can be work-related or personal. At least four types of commitment can be found in extant literature: affective, normative, continuance and instrumental commitment.

*Affective* commitment refers to a situation where a person truly believes in the focus of her commitment. This is indicated with phrases such as "I really want to do this". Affective commitment is the type of commitment that we typically refer to when we think of the construct. It is by definition a strong bond and thus difficult to influence from the outside.

*Normative* commitment refers to a situation where a person feels obliged to do something because of internal or external pressure. For this reason, in many cases, promises made in public are more binding than those that are kept to oneself.

*Continuance* commitment is the third type of commitment form. It is also known as escalation of commitment in the field of management. Continuance commitment refers to a situation where you have continued some activity for so long that the costs of aborting it are higher than those of completing the effort.

Finally, *instrumental* commitment is the most typical form of commitment and is often utilizing when motivating people to perform at a work place. The intent of the incentives is to tie the person to the commitment object (e.g. the objective of a project). [15]

Understanding how a person may be committed to a certain object is related to understanding what key concerns in that individual's work life. This can be modeled with a commitment net. A commitment net is web of concerns and their corresponding actions. It is a tool for making sense of what the priorities of an organization, a project, and an individual are. [15]

Literature [15] has established that a concern drives the behavior. In this study, we seek to understand the commitment of the developers when they were using the RESOLVEDD-strategy to better understand the results of their designs.

*2) Transparency*

In the ART model, Dignum [12] presents a rather narrow view of transparency, focusing on the transparency of the algorithms and data used, as well as their provenance and their dynamics. We argue that transparency has a more significant role in determining ethical design. As Turilli & Floridi [20] state, transparency acts as a pro-ethical circumstance that makes it possible to implement ethical principles into the design process.

The construct of transparency is used when referring to the visibility of information from the design and development process, as well as from the product itself. There are thus two types of transparency: transparency of systems, and transparency of systems development. The former refers to understanding how the systems are designed and why they act in certain ways in certain situations. The latter, on the other hand, refers to understanding what decisions were made during the development process, and why.

Transparency has been considered to be crucial for the ethical design and use of AI/AS since it provides a simple and objective way of understanding what an AI/AS is doing and why. Processes, products, values as well as design practices should be transparent in order to help to enhance human well-being and acceptance of technology [4, 12]. Without transparency in the actions of oneself or the system being developed, it is impossible to assess the justifications for the actions or the ethical principles behind them. E.g. if an autonomous vehicle crashes and we cannot understand why, ethical assessment of the incident and the decisions leading up to it is impossible as well. Systems need to be transparent so that the reasons behind unwanted results can be understood [4].

*3) Accountability*

To prevent misuse and to support EAD, accountability structures are needed [4]. In the ART model, accountability is seen as demand for the derivability of who accountable for the decisions made by system and its algorithms. In their more recent work, Dignum [12] defines accountability to refer to the explanation and justification of one's decisions and one's actions to the relevant stakeholders.

In order to consider someone accountable, there needs to be transparency in information, data, and design as discussed in the preceding sub-section. Therefore, transparency is required for accountability to be achievable. To achieve accountability, developers should be aware of the accountable matters that they are involved with and that are present in their systems.

In context of this study, accountability is used not only in the context of systems, but also in a more general sense. We consider, for example, how various accountability issues (legal, social) were taken into consideration during the design process.

*4) Responsibility*

Whereas accountability is related to the connection between one's decisions or actions and the stakeholders of the system, responsibility is an internal process. In order to act responsibly, one needs understand the meaning of their action. In the ART model, responsibility is related to idea of the chain of responsibility, even when there is no human agent as a direct cause of action there must be a linking chain to the responsible stakeholder. Therefore, artificial intelligence is an actor with a role in the chain of responsibility.

Responsibility in the context of this study connects the designer to the outside world, to others as stakeholders for example. In order to be responsible, one has to make weigh their own actions and to consciously evaluate their choices. E.g. one very simple way of considering responsibility would be to ask oneself "would I be fine with using my own system?".

*B. Study design*

The RESOLVEDD-strategy was empirically evaluated using a case study research method [21]. More specifically, we conducted case studies of five student projects that all utilized the RESOLVEDD-strategy. Yin [30] explains that the use of multiple case study makes it possible to have multiple data sources with rich in-depth investigations that would not be possible with a survey. This method also allowed the analysis within each case and across the cases to validate the observations by cross-referencing [21].

The study was conducted in an Information Systems (IS) course at the University of Jyväskylä. Bachelor level students were introduced to the RESOLVEDD-strategy as a part of the system design and development methods. In the course, the students were given the task of developing a concept and prototype of a futuristic innovation that could be possible in the near future, but which was not considered plausible with current technologies. The projects were carried out as a group work in five groups of 4-5 students. Choosing from a list., the students had to decide which technology they would want to utilize as part of their solution. For example, the students could make solutions that utilized Augmented Reality (AR), AI, or more specific technologies such as the Raspberry Pi computer.

### 3. FINDINGS

The findings from the analysis of the empirical data are reported here as topic-related Primary Empirical Conclusions (PEC). In total 5 PECs were formulated in the analysis. This section is structured into four sub-sections according to the research framework discussed in the preceding section.

*A. Commitment to Ethically Aligned Design*

All five teams had rather critical sentiments towards dealing with ethical issues or using ethical tool as a part of their product design. Using an ethical tool was perceived as something completely novel to them, and they did not seemingly place value on considering the ethical aspects on their project. This was despite of the fact that the employed method is focused on helping its users detect ethical issues. When considering commitment to EAD, it is important to understand what the true concerns of the developers are. In this case, the teams were more concerned about the usefulness and viability of their product than its ethical aspects.

PEC 1: While normative commitment to the use of Ethically Aligned Design brings immediate results, it will seize to exist when the external pressure is taken away. The RESOLVEDD-strategy needs adaptation in application context. In practice, group discussions were seen effective in addressing the ethical issues.

*B.* Transparency in design

Even though the teams were not affectively committed to using the ethical tool in their design process, they were required to follow the steps of the RESOLVEDD-strategy and to produce documents that increased the transparency and the visibility to the teams' decision-making process. Teams adapted the RESOLVEDD-strategy to fit their needs in order to carry out ethical thinking. The external pressure to use a specific method did not please the teams. Nonetheless, the necessitated use of the RESOLVEDD-strategy method did increase transparency and ensured that the ethical considerations of the teams were documented for later use. The teams remained skeptical, however, whether their documentation would be beneficial.

PEC2: When the RESOLVEDD-strategy is followed step-by-step a paper trail is born where each decisions made and the respective justification can be found. This produces transparency in the design process, but it does not promote transparency at the product layer.

*C.* Accountability in design

The question of accountability divided the teams. It was not clear to the teams who can be held accountable for the design. Teams defended their position (not being accountable) by arguing that the systems are only concepts and prototypes. They outsourced the issue of accountability to the end user, or they were not able to explain how it is managed from the legal or social viewpoints. The teams' lack of knowledge on accountability issues plays an important role.

PEC3: The RESOLVEDD-strategy does not deliver accountability.

*D.* Responsibility in design

Expecting the teams to engage in EAD and supporting their engagement in EAD by introducing an ethical tool made it possible to talk about the ethical issues related to their current projects. Our introduction to the RESOLVEDD-strategy could have been improved.

PEC 4: Requiring Ethically Aligned Design activated reflections on the developers' own sense of responsibility.

We also found that the teams were not keen on using the method, nor were they satisfied with the results they obtained by doing so. External pressure for the use of the tool nonetheless created tangible results, promoted EAD, and even supported the developers' sense of responsibility.

PEC 5: The mere presence of an ethical tool has an effect on ethical consideration creating more responsibility even when it the use of the method is not voluntary.

## 4. Discussion and Conclusions

In this study, we have evaluated empirically the RESOLVEDD-strategy for ethical decision-making through an exploratory, multiple case-study of five AI/AS projects. The study subjects were students and thus formed a limitation of the study that needs to be considered. We find that the limitation is not so relevant since Höst et al [22] finds that the differences between students and professionals is minor and not statistically significant. In fact, he recommends the use of students in software engineering studies. Runeson [23] finds similar improvement trends between undergraduate, graduate and professional study groups. For a novel topic in the field (such as EAD in our case), the students provide an excellent platform for an empirical evaluation, method development and experimentation. Future studies should consider case studies in industrial settings.

We found that while normative pressure to the use of Ethically Aligned Design brings immediate results, it will seize to exist when the external pressure is taken away (PEC1). RESOLVEDD increased transparency in the design process (PEC2) but it does not deliver accountability (PEC3). Requiring Ethically Aligned Design from the developers increased their sense of responsibility (PEC4). As a concluding finding it can be stated that the mere presence of an ethical tool has an effect on the ethical consideration exerted by developers, creating more responsibility even when the use of the method is not voluntary (PEC5).

The research framework formed in this study also has practical implications by making the level of ethically aligned design evaluable. We have shown, initially, that while it is possible to introduce EAD by force, results will not sustain over time. The RESOLVEDD-strategy needs to be adjusted in practice. One important adjustment done by our case teams was the introduction of group discussions as the primary means to do EAD in practice. Thus, a possible avenue for tailoring is to identify what are the practices that actually lead to favorable outcomes increasing transparency, responsibility and accountability.

## References

[1] C. Allen, W. Wallach and I. Smit, "Why Machine Ethics?" IEEE Intelligent Systems, vol. 21, (4), pp. 12-17, 2006 doi: 10.1109/MIS.2006.83.

[2] A. McNamara, J. Smith and E. Murphy-Hill, "Does ACM's code of ethics change ethical decision making in software development?", Proceedings of the 2018 26th ACM ESEC/FSE, pp. 729-733, 2018. doi:10.1145/3236024.3264833

[3] P. Abrahamsson and N. Iivari, "Commitment in software process improvement - in search of the process," Proceedings of the 35th HICSS, pp. 3239-3248, 2002. doi: 10.1109/HICSS.2002.994403.

[4] The IEEE Global Initiative on Ethics of Autonomous and Intelligent Systems. Ethically Aligned Design: A Vision for Prioritizing Human

Well-being with Autonomous and Intelligent Systems, First Edition. IEEE. 2019.

[5] T. Bynum, "Flourishing Ethics," Ethics and Information Technology, vol. 8, (4), pp. 157-173, 2006. doi: 10.1007/s10676-006-9107-1

[6] B. Friedman, "Value-sensitive Design," Interactions, vol. 3, (6), pp. 16-23, 1996. doi: 10.1145/242485.242493.

[7] B. Friedman, P. H. Kahn, A. Borning and A. Huldtgren, "Value Sensitive Design and Information Systems," in Early engagement and new technologies: Opening up the laboratory. Philosophy of Engineering and Technology, vol 16, N. Doorn et al. Eds. Dordrecht, Springer 2013. doi: 10.1007/978-94-007-7844-3

[8] J. Davis and L. P. Nathan, "Value sensitive design: Applications, adaptations, and critiques," in Handbook of Ethics, Values, and Technological Design, J. van den Hoven et al. Eds. Dordrecht, Springer 2015, pp. 11-40 doi: 10.1007/978-94-007-6970-0_3

[9] A. Wynsberghe, "Designing Robots for Care: Care Centered Value-Sensitive Design," Sci. Eng. Ethics, vol. 19, (2), pp. 407-433, 2013. doi: 10.1007/s11948-011-9343-6

[10] J. Miller, B. Friedman and G. Jancke, "Value tensions in design: The value sensitive design, development, and appropriation of a corporation's," Proceedings of the ACM Group 2007, pp. 281-290, 2007. doi: 10.1145/1316624.1316668.

[11] J. Bryson and A. Winfield, "Standardizing Ethical Design for Artificial Intelligence and Autonomous Systems," Computer, vol. 50, (5), pp. 116-119, 2017. doi: 10.1109/MC.2017.154

[12] V. Dignum, "Responsible autonomy," arXiv Preprint arXiv:1706.02513, 2017.

[13] A. W. Flores, K. Bechtel and C. T. Lowenkamp, "False positives, false negatives, and false analyses: a rejoinder to "Machine bias: there's software used across the country to predict future criminals, and it's biased against blacks"," Federal Probation, vol. 80, (2), pp. 38, 2016.

[14] D. Graziotin, X. Wang and P. Abrahamsson, "Are happy developers more productive?" in Product-Focused Software Process Improvement, pp. 50-64, 2013. doi: 10.1007/978-3-642-39259-7_7

[15] P. Abrahamsson, "Commitment Nets in Software Process Improvement," Annals of Software Engineering, vol. 14, (1), pp. 407-438, 2002. doi: 1020526329708".

[16] R. S. Pfeiffer and R. P. Forsberg, Ethics on the Job: Cases and Strategies. Wadsworth Publishing Company, 1993.

[17] C. Johansen, "Teaching the ethics of biology," The American Biology Teacher, vol. 62, (5), pp. 352-358, 2000.

[18] N. E. Bowie, "A Kantian Theory of Meaningful Work," J. Bus. Ethics, vol. 17, (9), pp. 1083-1092, 1998.

[19] P. Abrahamsson, "Rethinking the Concept of Commitment in Software Process Improvement," Scandinavian Journal of Information Systems, vol. 13, (1), 2001.

[20] M. Turilli and L. Floridi, "The ethics of information transparency," Ethics and Information Technology, vol. 11, (2), pp. 105-112, 2009. doi: 10.1007/s10676-009-9187-9

[21] R. K. Yin, Qualitative Research from Start to Finish, Second edition. New York, Guilford Press, 2016.

[22] M. Höst, B. Regnell and C. Wohlin, "Using Students as Subjects A Comparative Study of Students and Professionals in Lead-Time Impact Assessment," Empirical Software Engineering, vol. 5, (3), pp. 201-214, 2000. doi 1026586415054".

[23] P. Runeson, "Using students as experiment subjects – an analysis on graduate and freshmen student data," Proceedings of the 7th International Conference on EASE. pp. 95-102 2003.